\documentstyle[prl,twocolumn,float,epsf,aps,graphicx]{revtex}
\begin{document}
\draft
\title{AC DRIVEN JUMP DISTRIBUTIONS ON A PERIODIC SUBSTRATE}
\author{M. Borromeo}
\address{Dipartimento di Fisica, and Istituto Nazionale di Fisica Nucleare,
Universit\`a di Perugia, I-06123 Perugia (Italy)}
\author{F. Marchesoni}
\address{Istituto Nazionale di Fisica della Materia, Universit\`a di Camerino, 
I-62032 Camerino (Italy)}
\date{Received: \today}
\maketitle
\begin{abstract} A driven Brownian particle (e.g. an adatom on a surface)
diffusing on a low-viscosity, periodic substrate may execute multiple jumps. In the presence
of an additional periodic drive, the jump lengths and time durations
get statistically modulated according to a synchronization mechanism reminiscent
of asymmetric stochastic resonance. Here, too, bistability plays a key role,
but in a dynamical sense, inasmuch as the particle switches between
 locked and running states.
\end{abstract}

~


~

Corresponding author: Dipartimento di Fisica, Universit\`a di Perugia,
I-06123 Perugia (Italy); Tel: +39 075 5853033; Fax: +39 075 44666; E-mail:
marchesoni@pg.infn.it

~


The diffusion of a forced Brownian particle on one- or two-dimensional periodic
substrates provides an archetypal model of relaxation in condensed phase with application
to adsorbates on crystal surfaces \cite{1}, transport in superionic conductors
\cite{2}, acoustoelectric currents in one-dimensional channels \cite{3},
dispersion of particles in optical traps \cite{4}, or rotation of molecules
in solids \cite{5}. The diffusive process from this category best studied in 
the current literature \cite{6}-\cite{16} is the Brownian motion in a one-dimensional
washboard potential 
\begin{equation}
\label{1}
\ddot x= -\gamma \dot x -\omega_0^2 \sin x + F +\xi(t)
\end{equation}
(in rescaled units),  where the force terms on the r.h.s. represent,
 respectively, a viscous 
damping with constant $\gamma$, a spatially periodic, tilted substrate 
modeled by the potential
\begin{equation}
\label{2}
V(x, F) = \omega_0^2 (1-\cos x) - Fx,
\end{equation}
and a stationary Gaussian noise with zero mean $\langle \xi(t) \rangle = 0$
and autocorrelation 
function $\langle \xi(t) \xi(0) \rangle = 2\gamma kT \delta(t)$. \\

\begin{figure}[h]
\includegraphics[width=8cm, height=6cm]{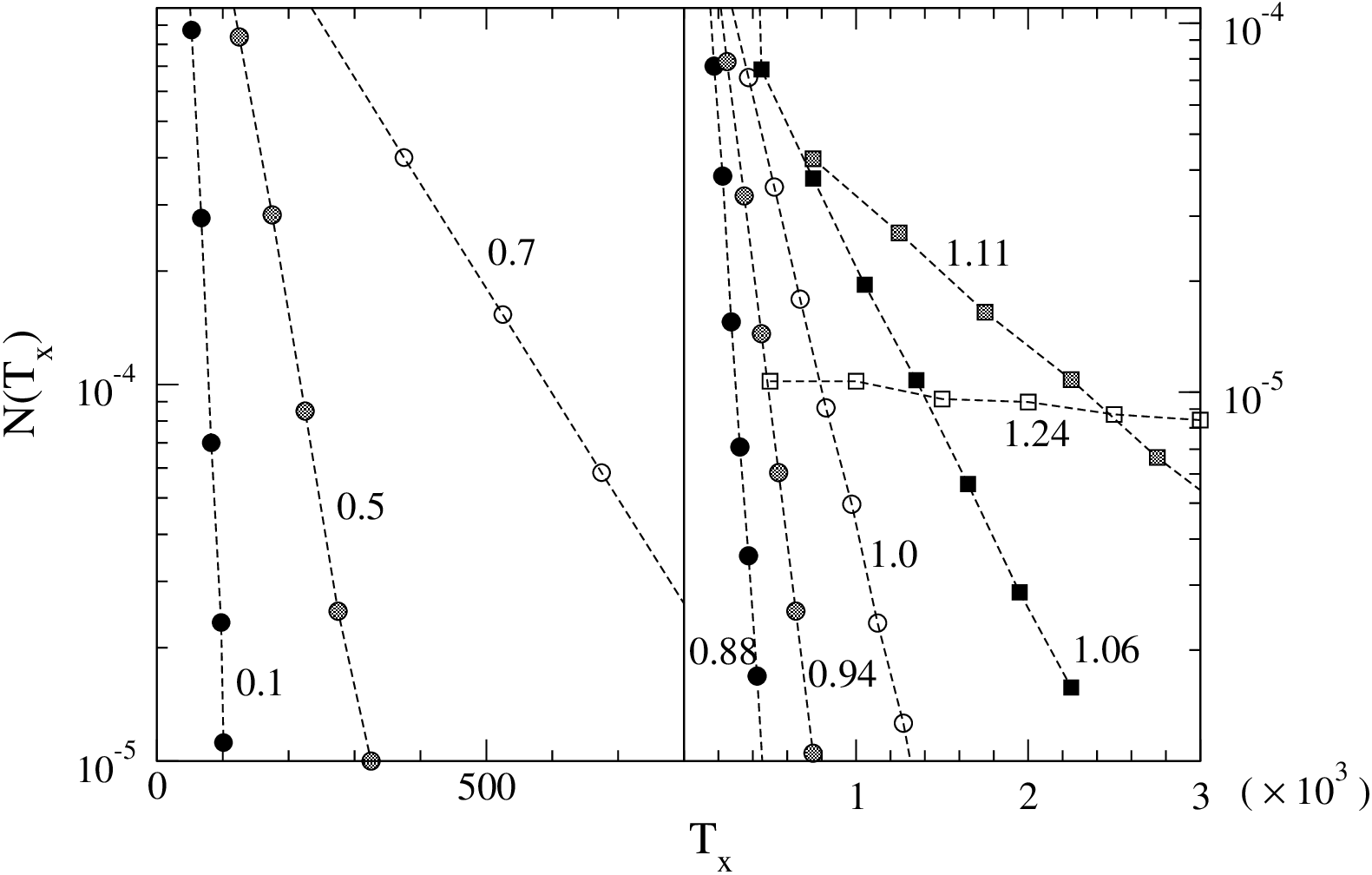}
\caption{\label{F1} Jump distributions in the absence of modulation, 
$\Delta F=0$, for different values of the dc drive $F_0$ below (a) and in 
the vicinity of the threshold $F_2$ (b). $F_0$ is expressed in units of
$F_2$ with $F_2=0.085$.
 Other parameter values:
$\gamma/\omega_0 = 0.03$, $kT/\omega_0^2=0.3$.
   }
\end{figure}

In the {\it low damping} limit, $\gamma \ll \omega_0$, (strictly speaking
it would suffice that $\gamma < 1.2 \omega_0$ \cite{6}) the Brownian 
particle executes multiple jumps, namely over many a lattice spacing $a=2\pi$,
both in the forward and in the backward direction \cite{15,16}; with increasing $F$
the length- and time-distributions of such multiple jumps, denoted,
 respectively, by $N(X)$ and $N(T_x)$, change from an almost exponential curve
(locked regime, $F<F_1 \equiv (4/\pi)\gamma \omega_0$ \cite{7})
to a kind of bimodal curve with a slow exponential tail (running regime
$F>F_2 \simeq 3.36 \gamma \omega_0$ \cite{9,12}, see also Fig.
\ref{F1}). To a good approximation $N(X)$ and $N(T_x)$ are related by the 
scaling law $X=(F/\gamma)T_x$ \cite{14}, where $F/\gamma$ is the average velocity
of the particle in the running regime \cite{6}.\\

In a recent paper Talkner {\it et al.} \cite{17} addressed the effects of a weak
periodic drive on the jump statistics in the intermediate regime with
$kT\ll F\ll \omega_0^2$
(exponential hopping regime \cite{13}):
An enhancement of the multiple jumping probability has been predicted in the adiabatic
limit of very low modulation frequencies. Taking on their challenge to a
full investigation, we have simulated numerically the Langevin equation (LE)
(\ref{1}) with the dependent drive
\begin {equation}
\label{3}
F \rightarrow F(t)=F_0 + \Delta F \cos(\Omega t),
\end{equation}
$\Delta F>0$. Note that the time origin has been set to zero without loss
of generality and the modulation amplitude $\Delta F$ has been kept small
compared to the dc term $F_0$ in all simulation runs.\\

In the present report we focus on the time distributions $N(T_x)$ of the 
multiple jumps at the crossover between the locked and the 
running regime: A significant synchronization mechanism takes place reminiscent 
of stochastic resonance (SR) in a bistable system \cite{18}.\\
\bigskip
\medskip

At low temperatures $kT \ll \omega_0^2$, and in the exponential hopping regime,
$\gamma \ll \omega_0^2$ and $F_0 < F_1$, jumps are thermally activated and, therefore, relatively short;
their direction (either forwards or backwards) is controlled by the intensity
of the dc drive $F_0$ relative to $F_1$ \cite{15,16}. Under such conditions, the 
Brownian particle sits in a {\it locked} state with low mobility, namely
$\gamma \mu \equiv \gamma \langle \dot x \rangle/F_0 \ll 1$. On raising $F_0$
above the threshold value $F_2 \simeq 3.36 \gamma \omega_0$ ($F_2$ 
depends weakly on the temperature) $\gamma \mu$ jumps abruptly up to close to unity,
thus signalling that the particle has switched into the {\it running} state
with average velocity $\langle \dot x \rangle =F_0/\gamma$. As the particle
flies over the washboard potential it executes sporadic, short stops at the 
bottom of some potential well. For numerical purposes the diffusing particle is
deemed as locked (or retrapped) any time it sojourns within two adjacent 
$V(x,F)$ barriers for a time interval longer than $(2\gamma)^{-1}$ -- the
characteristic relaxation time of the energy variable. In Fig. \ref{F1} we display
the normalized distributions of the jump time durations, $N(T_x)$, for different 
values of $F_0$, ranging from the locked, well into the 
running regime. As expected, all of our distributions exhibit exponential
tails with time constant $\bar T_x$, termed from now on the average jump duration,
that increases exponentially with $F_0$ \cite{7,15}. No dramatic change in 
the $N(T_x)$ profile occurs in the vicinity of the thresholds $F_1$ and $F_2$.\\

\begin{figure}[h]
\includegraphics[width=8cm, height=6cm]{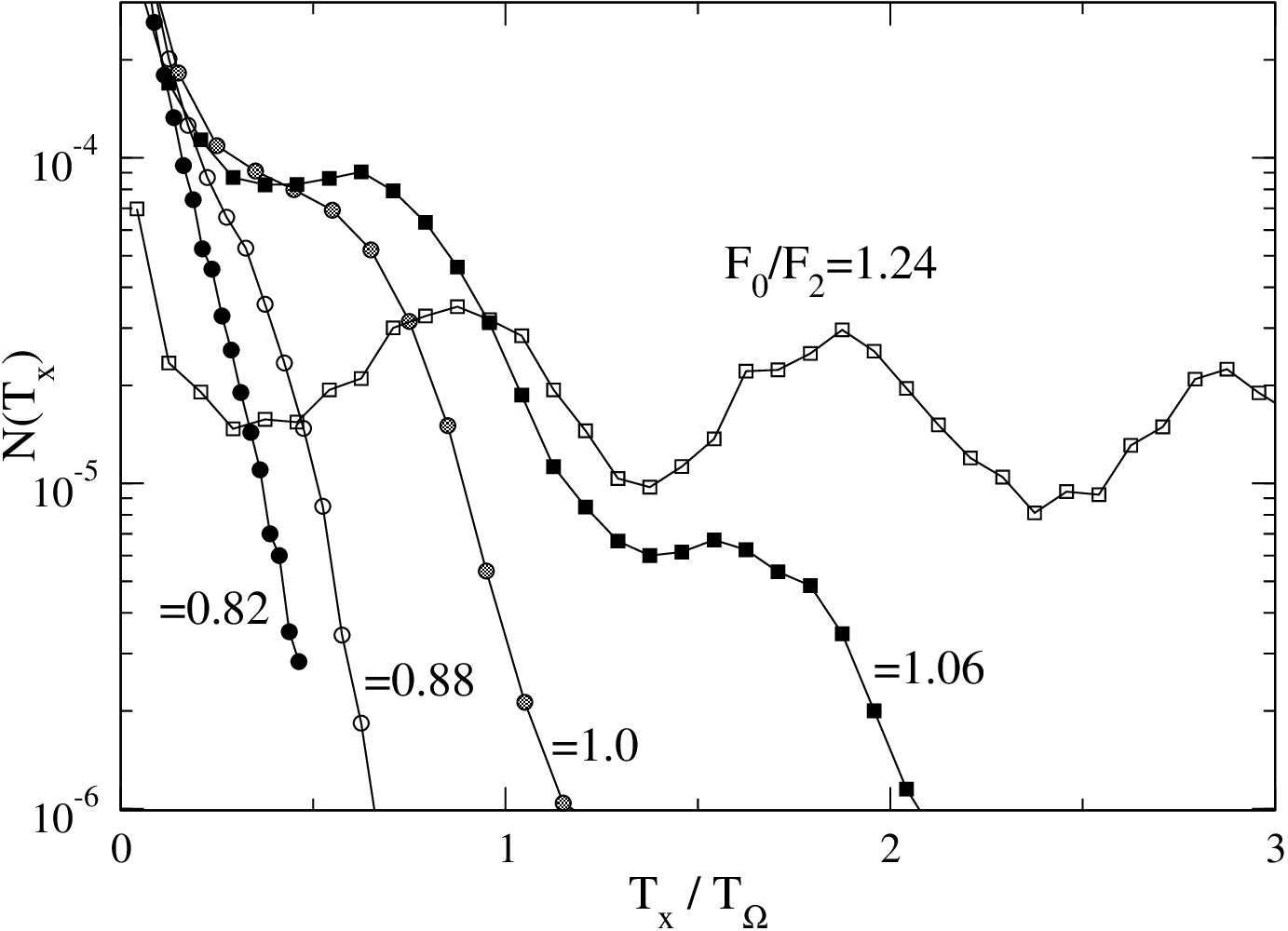}
\caption{\label{F2} Jump distributions in the presence of an ac drive with
constant amplitude, $\Delta F/F_2=0.12$, and period, $T_{\Omega}=2\pi/\Omega =6.4\cdot 10^4$, 
and for
different values of the static tilt $F_0$. 
Other parameter values are as in Fig. 1.
   }
\end{figure}

The effect of the weak periodic modulation (\ref{3}) on the jump statistics is
illustrated in Fig. \ref{F2}, where $N(T_x)$ is plotted for the same ac drive,
but increasing values of $F_0$ (across the threshold $F_2$).
For $F_0<F_2$ the distribution of the jump durations is well reproduced by an exponential
function; for
$F_0 \sim F_2$ the curve bends to form a broad hump around $T_x=T_{\Omega}/2$
(here $T_{\Omega}=2\pi/\Omega$ is the modulation period); finally, for 
$F_0 > F_2$ the distribution develops a full-blown peak structure, with maxima
centered at $T_n = nT_{\Omega}$, with $n=1,2, \dots$ superimposed on a slowly decaying
exponential tail. On a closer look one recognizes that even for $F_0=F_2$ the jump distribution
is multi-peaked; however, the first peak dominates apparently over the others,
thus indicating a strong drive-jump synchronization \cite{19}.
More remarkably, the positions of the $N(T_x)$ maxima shift from 
$T_n=(n-1/2)T_{\Omega}$ at $F_0 \simeq F_2$ to $nT_{\Omega}$ for $F_0 \gg F_2$
\cite{20}.\\

The qualitative interpretation of these results is simple. For $F_0 <F_2$
the length and, therefore, the time duration of the multiple jumps is relatively
short, -- we recall that $kT \ll \omega_0^2$, -- anyway, shorter that the modulation period
$T_{\Omega}$. We are dealing with an adiabatic situation, where the slowly
varying external parameter $F(t)$ during the jump process can be regarded as
constant; the ensuing distribution $N(T_x;t)$ is expectedly exponential 
\cite{12} at any times, so that $N(T_x)=\langle N(T_x,t)\rangle$, $\langle
\dots \rangle$ denoting the average over a modulation cycle. To this purpose,
it should be remembered  that the jump rate, namely the number of jumps
per unit of time, is maximum at $F_0 \simeq F_2$ and decays exponentially with
$F_0$ above threshold \cite{15}. For $F_0 \simeq F_2$ the system operates at
the unlocking threshold; a small ac drive amplitude $\Delta F$ suffices to
 modulate the system, alternately, from the locked to the running state and vice versa.
No surprise that most jumps last the order of one half forcing period, hence the appearance 
of a bump in the $T_x$ distribution at $T_{\Omega}/2$. For $F_0 > F_2$ the 
Brownian particle is in the running state most of the time, so that only infrequent, very long, 
jumps may take place.
Clearly, chances are that rare relocking events may happen mostly 
as $F(t)$ hits a minimum $F_0 - \Delta F$. As a consequence, we expect that $T_x$
are kind of "multiple" of $T_{\Omega}$ (jump-drive phase-locking).
 Notice that at the threshold the average
residence time in the locked state $\bar T_r$ equals the running time $\bar T_x$,
whereas in the running regime $\bar T_r$ becomes negligible \cite{7}.\\

\begin{figure}[h]
\includegraphics[width=8cm, height=6cm]{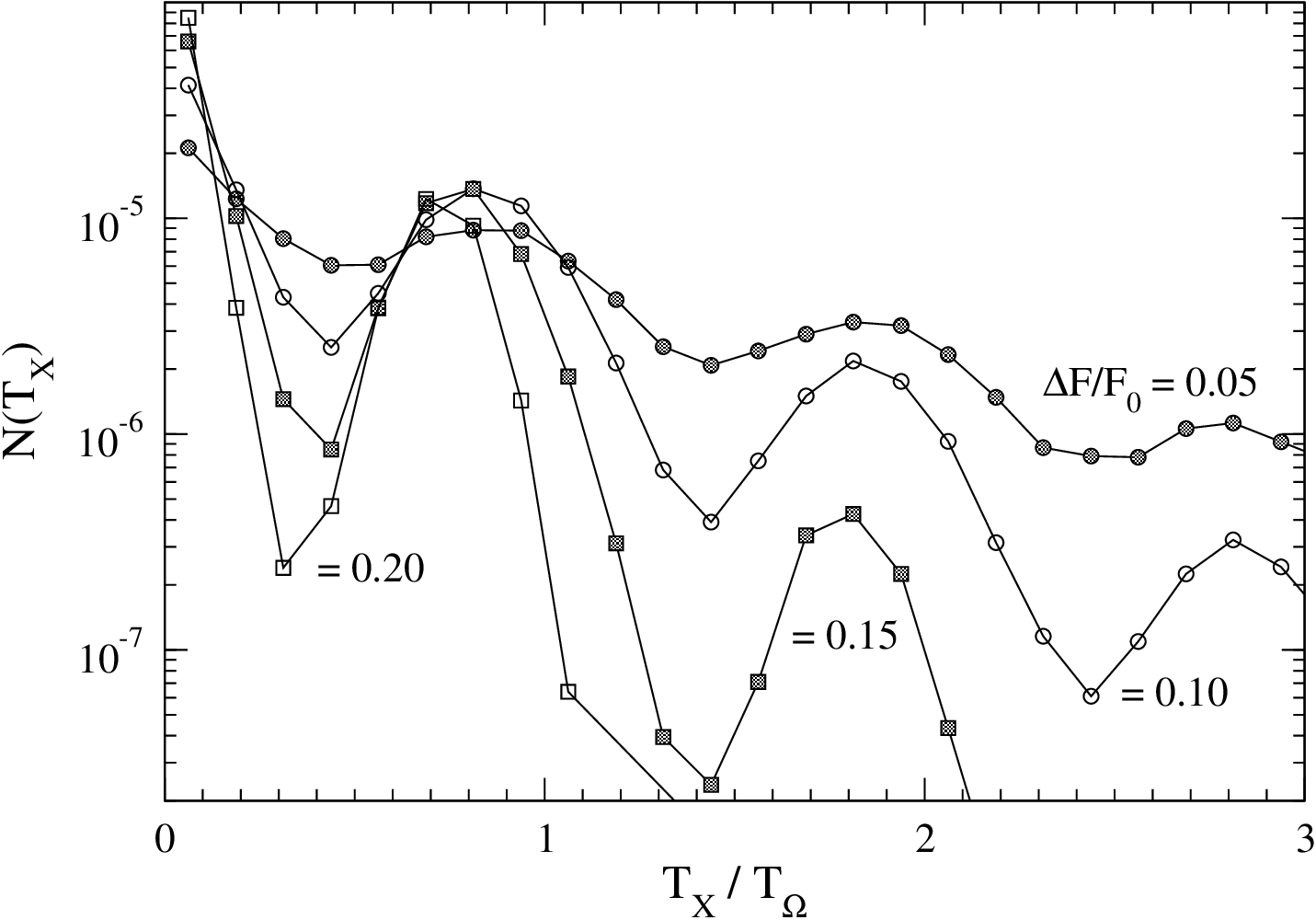}
\caption{\label{F3}
Dependence of the jump distributions on the modulation amplitude $\Delta F$
at constant period, $T_{\Omega}=8\cdot 10^3$, and static tilt, $F_0/F_2=1.24$.
Other parameter values are as in Fig. 1.\\
   }
\end{figure}

The dependence of the jump synchronization mechanism on the amplitude $\Delta F$
of the ac drive is displayed in Fig. \ref{F3}.
 After subtracting the exponential background due to the random 
jumps, we concluded that in the weak drive regime, $\Delta F \ll
F_0$,
the fraction of the synchronized jumps (of any duration
$T_n$) grows quadratically with $\Delta F$. As a matter of fact, changing the
ac drive amplitude affects the background decay constant, as well. At finite
$\Delta F$ values the background seems to fall off faster than in the absence
of modulation. This suggests that increasing $\Delta F$ has a twofold effect:
It modulates more and more effectively the jump lengths and, simultaneously,
diminishes the effective dc drive ($\bar F_0$, to be defined in the following)
below the input value $F_0$
by an amount that depends on the forcing period.\\

\begin{figure}[h]
\includegraphics[width=8cm, height=6cm]{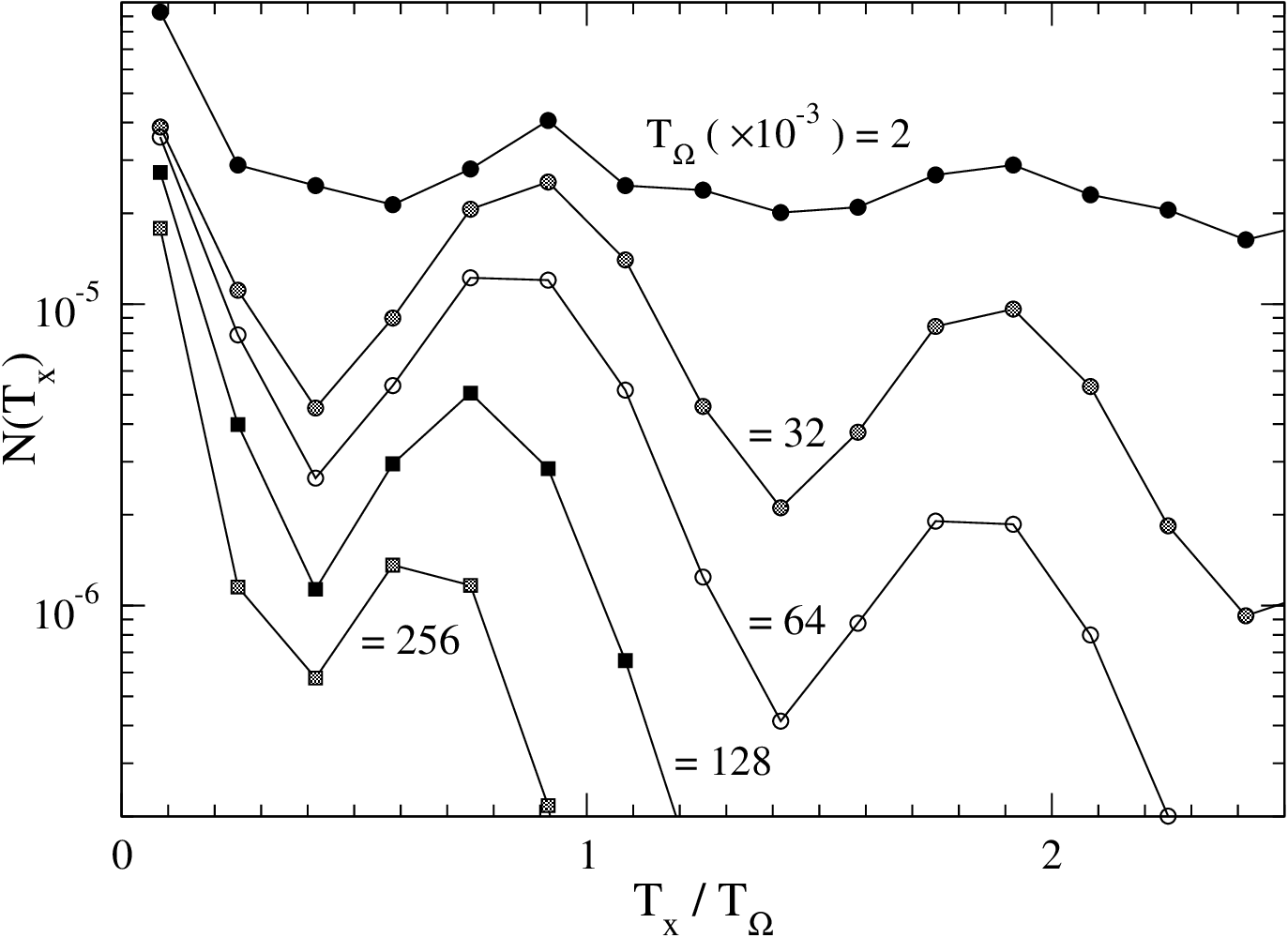}
\caption{\label{F4} 
Dependence of the jump distributions on the modulation period
$T_{\Omega}$ (in units of $10^3$)
at constant amplitude, $\Delta F/F_2=0.12$, and static tilt, $F_0/F_2=1.24$.
Other parameter values are as in Fig. 1.
   }
\end{figure}

Finally, in Fig. \ref{F4} we show the dependence of the jump distribution
$N(T_x)$ on the modulation period. Note that all distributions plotted here 
have been normalized to unity after rescaling $T_x$ to $T_x/T_{\Omega}$.
The overall $\Omega$ dependence of the synchronization mechanism is apparent:
For short modulation periods the jump-drive phase-locking is hardly
detectable; on increasing $T_{\Omega}$ the $N(T_x)$ peak structure reaches first
its maximum and, eventually, merges into the background for exceedingly long
forcing periods. The qualitative interpretation of these results is also
straightforward. The case of large $T_{\Omega}$ corresponds to the adiabatic 
situation illustrated above, when we discussed the distributions of Fig.
\ref{F2} with $F_0 <F_2$; the case of small $T_{\Omega}$ values represents
the opposite situation, where the external time-dependent parameter $F(t)$
varies so fast in time, that the diffusing particle only responds to its effective
average value $\bar F_0$ insensitive to its temporal modulation
-- technically speaking, also an adiabatic limit; in the intermediate regime $T_{\Omega}
\simeq \bar T_x$ the synchronization mechanism becomes more efficient, due to the
interplay between externally driven and thermally activated 
 locked-to-running transitions \cite{18}. The apparent increase of the background
steepness with $T_{\Omega}$ is a graphic artifact introduced by rescaling 
the units of $T_x$; when plotted versus 
molecular dynamics time-units all curves of Fig. \ref{F4} 
decay with time constants corresponding to 
an effective tilt $\bar F_0$ such that $F_0 - \Delta F < \bar F_0 <F_0$.\\

We attempt now at interpreting the numerical results reported so far as a 
manifestation of {\it stochastic resonance} (SR), -- an intriguing phenomenon
of nonlinear physics, where an optimal amount of noise has the capability
of enhancing the rate of synchronous switches between the local minima of a
(possibly, asymmetric) bistable system driven by a weak periodic modulation.
Stochastic resonance rests upon a noise controlled phase-locking mechanism
which can be vividly illustrated in term of switch-time distributions \cite{19}.
A convenient choice is provided by the distributions $N_1(T_{\pm})$ of the residence times
$T_{\pm}$ of the system in either state (denoted by $\pm$, respectively).
As the modulation forces the switch events by perturbing the symmetry of the
system, $N_1(T_{\pm})$ peak respectively at $T_n = T^{(0)}_{\pm} + nT_{\Omega}$, 
with $n=1,2, \dots$ and $T^{(0)}_{+}+T^{(0)}_{-}=T_{\Omega}$. For a symmetric system,
$T^{(0)}_{\pm}=T_{\Omega}/2$. An optimal input-output synchronization sets in
when the first peak of  $N_1(T_{\pm})$ at  $T^{(0)}_{\pm}$ dominates over the remaining peaks and the exponential 
random switch background \cite{20}.\\

In the system at hand all ingredients of the synchronization-based 
(or {\it bona-fide}) SR \cite{19} are easily recognizable: (i) {\it bistability} occurs at the
threshold $F_2$, where the particle mobility switches between two dynamical 
states, namely the locked state with $\gamma \mu=0$ and the running state with
$\gamma \mu =1$. The residence times in the two states are denoted here by
$T_r$ and $T_x$, respectively; (ii) the {\it noise} control parameter is the
equilibrium temperature $T$. Note that SR may occur even at low damping \cite{18};
(iii) the external {\it periodic} drive, $\Delta F \cos(\Omega t)$, Eq. (\ref{3}),
 tilts the system towards either dynamically stable state, alternately, thus
determining the multi-peaked $T_x$ distributions of Figs. \ref{F2}-\ref{F4}.\\

A few well-known SR features are apparent: (1) The $N(T_x)$ distributions develop their first
(and strongest) peak for $T_{\Omega} \simeq 2\bar T_x$ (see Figs. \ref{F2} and
\ref{F4}); (2) The first peak, at $T_x = T^{(0)}_x$, shifts from close to
$T_{\Omega}/2$ in the symmetric case, $F_0=F_2$, 
to $T_{\Omega}$ in the running regime,
$F_0 > F_2 + \Delta F$. This is a typical asymmetry effect \cite{20}, as
on raising $F_0$ above $F_2$, $\bar T_r$ gets smaller than $\bar T_x$;
(3) The peak structure grows sharper on increasing $\Delta F$, according to a
 characteristic SR quadratic law \cite{18}.\\

Peculiar to our model are the properties following: (4) On increasing $F_0$ 
at $\Delta F$, $T$ and $\Omega$ constant, more and more peaks shoot up from the background,
thus reducing the input-output synchronization (see Fig. \ref{F2}). The
explanation in terms of the SR phenomenology is simple: The jump times $T_x$ 
coincide with the residence times in the running state, which for
 $F_0>F_2$ gets more stable than the locked state; accordingly, $\bar T_x$
grows much longer than $T_{\Omega}$ or, equivalently, relative to the directed
running-to-locked transition, the noise intensity is too low for SR 
to take place; hence the multi-peaked structures plotted in Figs. \ref{F2}-
\ref{F4} \cite{18,21}; (5) Contrary to most SR studies, here the background
constant depends on the ac drive, namely on $\Delta F$ and, though to a smaller extent,
on $T_{\Omega}$. A pictorial explanation of such a property is provided
by Fig. 11.24 of Ref. \cite{6}. Consider the most symmetric case with $F_0=F_2$;
adding $\Delta F$ to $F_2$ makes the running state more stable than the locked
state and, vice versa, subtracting $\Delta F$ from $F_2$ makes the running state 
less stable; however, the imbalance toward the locked state at 
$F_0=F_2-\Delta F$ is not compensated by the imbalance towards the running state
at $F_0=F_2+\Delta F$; therefore, an ac drive contributes to the effective
 static tilt $\bar F_0$, with $\bar F_0<
F_2$, defined implicitly through the average jump duration as
$\bar T_x(F_0, \Delta F, \Omega) \equiv \bar T_x(\bar F_0)$.\\

\newpage
\vskip 2 cm

\vskip 2 cm

\vskip 2 cm

\vskip 2 cm
\end{document}